\documentstyle[12pt]{article}
\title{Differential equations to compute $\hbar$
corrections of the trace formula}
\author{G\'abor Vattay\\
E\"otv\"os University, Department of Solid State Physics\\
H-1088 Budapest, M\'uzeum krt. 6-8, Hungary }

\date{\today}
\begin{document}
\maketitle

\begin{abstract}
In this paper a new method for computation of higher order
corrections to the saddle point approximation of the
Feynman path integral is introduced. The saddle point approximation
leads to local Schr\"odinger problems around classical orbits.
Especially, the saddle point approximation leads to
Schr\"odinger problems around classical periodic orbits
when it is applied to the trace of Green's function.
These local Schr\"odinger problems, in semiclassical
approximation, can be solved exactly on the basis of local analytic
functions. Then the corrections of the semiclassical result can be treated
perturbatively. The strength of the perturbation is
proportional to $\hbar$. The perturbation problem leads
to ordinary differential equations. We propose these
equations for numerical calculation of corrections, since
they can easily be solved by computers. We give quantum
mechanical generalizations of the semiclassical zeta functions,
spectral determinant and trace formula.
\end{abstract}

Feynman's path integral in quantum mechanics\cite{Feynman} and similar
path integrals for stochastic systems are the most intuitive tools of
modern theoretical physics. But calculations with path integral
are difficult\cite{Schulman} in general. One can often find numerical solution
of the underlying partial differential equation easier. The most convenient
asymptotic method to evaluate the path integral is the saddle point
approximation. The leading gaussian approximation is easy to perform
and gives very good results. This is the case when we calculate the
trace of the Green's function of the one-body Schr\"odinger
equation in gaussian approximation and get the
Gutzwiller trace formula\cite{Gutzwiller}. From the
Gutzwiller trace formula one can recover the eigenenergies of
the Schr\"odinger equation for bound systems\cite{Chaos} and the resonances
of the
S matrix for scattering systems in semiclassical approximation\cite{Rice}.
There are many attempts to improve the semiclassical approximation to
get more accurate energies and resonances. But the gaussian approximation
has its inherent limitation and one should go beyond it to improve the
accuracy.
Recently, Gaspard et al.\cite{Gaspard} computed corrections of the Gutzwiller
trace formula and showed that the resonances of the two and three disk
scattering systems\cite{Rice,Wirzba} can be improved considerably. They have
used the
usual Feynman graph technique of the perturbation theory and
computed large number of graphs to get the corrections.
In general the conventional graph calculus is
very cumbersome. We give here two simple examples, where the corrections
can be computed and an agreement with the calculations in
Ref.\cite{Gaspard} can be checked.

In this paper
we show how the corrections to the gaussian approximation can be
computed using simple ordinary differential equations.
We hope the
results can be generalized for the field theory and for many body systems
in the near future.
Here we treat the one body non-relativistic Schr\"odinger equation and give
the $\hbar$ corrections of
the Gutzwiller trace formula and the spectral determinant introduced
by Voros\cite{Voros}.

\section{Saddle point evaluation of the trace}

The time domain propagator $G(q,q',t)$ of the Schr\"odinger equation
is defined by
\begin{equation}
i\hbar \partial_t G(q,q',t) -\hat{H} G(q,q',t)=\delta(q-q')\delta(t),
\end{equation}
where $$\hat{H}=-\frac{\hbar^2}{2}\Delta+U(q)$$
is the Hamilton operator in units of $m=1$.
The path integral representation of the propagator
\begin{equation}
G(q,q',t)=\int {\cal D}q'' e^{\frac{i}{\hbar} S(q,q',t | q'')},
\end{equation}
where $\int {\cal D} q''$ represents the functional integral measure
for all the paths connecting $q$ with $q'$ in time $t$, and
$S(q,q',t|q'')$ is the classical action between $q$ and $q'$
computed along a given path $q''$.
The resolvent of the Hamilton operator, the energy domain Green's function,
can be recovered as the Fourier transform of the propagator
\begin{equation}
G(q,q',E)=\frac{1}{i\hbar}\int_0^{\infty}dt \exp\left(\frac{i}{\hbar} E t
\right)G(q,q',t).
\end{equation}
The trace of time and energy domain Green's functions
can be expressed with the help of the eigenenergies of the Schr\"odinger
equation:
\begin{eqnarray}
\mbox{Tr} G(t)& =&\sum_{n=0}^{\infty} \exp\left(-\frac{i}{\hbar} E_n t\right),
\\
\mbox{Tr} G(E) & =&\sum_{n=0}^{\infty} \frac{1}{E-E_n}.
\end{eqnarray}
The path integral expression for the trace of the propagator is
\begin{equation}
\mbox{Tr} G(t)=\int dq G(q,q,t)=\int {\cal D}q'' e^{\frac{i}{\hbar} S(t|q'')},
\end{equation}
where $\int {\cal D}q''$ now represents the functional integration
for closed paths.
The spectral determinant $\Delta(E)=\det(E-\hat{H})=\prod_n(E-E_n)$,
whose zeroes are the eigenenergies or resonances, is related to the
trace of energy domain Green's function through the
logarithmic derivative
\begin{equation}
\mbox{Tr} G(E)=\frac{d}{dE} \log \Delta(E)\label{7}.
\end{equation}
In the saddle point approximation we assume that the
leading contribution to the path integral is coming from the
neighbourhood of paths for which the classical action is stationary.
This condition singles out the classical trajectories from the infinite
variety of possible paths. The trace of the propagator can be written
as a sum for the classical periodic orbits:
\begin{equation}
\mbox{Tr} G(t)=\sum_{\mbox{p.o.}}\int {\cal D} q_p \exp\left(\frac{i}{\hbar}
W_p(x,t)\right),
\label{trg}
\end{equation}
where $\sum_{\mbox{p.o.}}$ denotes the summation for the classical
primitive periodic orbits, $\int {\cal D} q_p$ denotes a functional integral
in the neighborhood of periodic orbits and $W_p(x,t)=
\sum_{n} w_n(x-q(t))^n/n! $ denotes
the classical action expanded in multi-dimensional power series around the
periodic orbit.
The energy domain Green's function can be recovered by Fourier transformation.

Here we do not investigate the general validity of saddle point
approximation.
However, it is important to note that the regions around
periodic orbits, where the action is expanded in power series, may
overlap and this causes some overcounting in the formula (\ref{trg}).
Therefore in computations the number of periodic orbits included in
the sum should depend on the order of truncation of the power series.
In the semiclassical or gaussian approximation the method proposed
by Berry and Keating\cite{Keating} can deal with this
problem. We hope, that their method can be extended for
the situation discussed here.

Now, the traditional computation goes as follows:
In the gaussian approximation the power series expansion is truncated at
quadratic terms and the Gauss integrals are evaluated. This way one
can derive Gutzwiller's trace formula. Corrections of the gaussian
approximation can be found by expanding the action to higher orders,
expanding the exponential and performing the gaussian cumulant integrals.

However, we have a better choice at this point. We can realize that
the terms in (\ref{trg}) are similar to the original
one in the neighborhood of periodic orbits except
the classical actions are given in power series form.
We can look for
the partial differential equations corresponding to these local path integrals.
These partial differential equations are Schr\"odinger equations in
a power series form around periodic orbits in the configuration space.
The terms in (\ref{trg})
are traces of the propagators $\mbox{Tr} G_p(t)$
of these  local Schr\"odinger problems
\begin{equation}
\mbox{Tr} G(t)=\sum_{\mbox{p.o.}} \mbox{Tr} G_p(t).
\end{equation}
By Fourier transformation we can also get the energy domain trace term by term
\begin{equation}
\mbox{Tr} G(E)=\sum_{\mbox{p.o.}} \mbox{Tr} G_p(E).
\end{equation}
The Gutzwiller-Voros spectral determinant is then a product of
the spectral determinants of local Sch\"odinger problems
\begin{equation}
\Delta(E)=\prod_{\mbox{p.o.}} \Delta_p(E),\label{prod}
\end{equation}
which are related to the traces as logarithmic derivatives
\begin{equation}
\mbox{Tr} G_p(E)=\frac{d}{dE} \log \Delta_p(E).
\end{equation}
The product (\ref{prod}) is the direct generalization of the
Selberg type product\cite{Selberg,Voros} as we will see later.

In general, there are many different types of closed periodic
orbits which can contribute to the product (\ref{prod}). Some of them are
zero length or equ\-i\-lib\-ri\-um\cite{Gaspard} orbits which can be well
treated by conventional methods. The spectral determinant of the zero
length orbits gives a smooth contribution, which is the counterpart of the
Weyl or Thomas-Fermi terms. From now on we neglect these terms
since they do not change the location of the zeroes of the
spectral determinant. Also, the periodic orbits in the complexified
phase space of the hamiltonian system can occur\cite{Berry1}.
Special contributions can come from diffraction cycles introduced
recently\cite{Buslaev,Wirzba,Vattay3,Vattay4}. Here we
concentrate on usual classical periodic orbits and we
discuss other terms elsewhere.

\section{Local spectra of the Schr\"odinger equation}

To compute the local spectral determinants $\Delta_p(E)$ we have to solve the
local Schr\"odinger
problem in the neighborhood of a classical periodic orbit.
The most convenient way to do this is to rewrite the Schr\"odinger
equation
\begin{equation}
i\hbar \partial_t \psi =-\frac{\hbar^2}{2m} \Delta \psi + U\psi
\end{equation}
with the usual ansatz
\begin{equation}
\psi=\varphi e^{iS/\hbar},\label{ansatz}
\end{equation}
which yields the following equation
\begin{equation}
-\varphi \partial_t S +i\hbar\partial_t\varphi=-\frac{\hbar^2}{2}
\left( \Delta\varphi+2i/\hbar\nabla\varphi\nabla S+i/\hbar\varphi\Delta
S -1/\hbar^2 \varphi (\nabla S)^2\right) + U\varphi.
\end{equation}
Here we have many possibilities to group the terms since we
have not made any restriction for $S$ and $\varphi$.
Our main concern is to separate the classical and the quantum
time evolution. Therefore, we require the phase to
fulfill the Hamilton-Jacobi equation
\begin{equation}
\partial_t S +\frac{1}{2}(\nabla S)^2 + U=0.
\end{equation}
At a given phase $S(x,t)$ the complex amplitude fulfills
\begin{equation}
\partial_t\varphi+\nabla\varphi\nabla S+
\frac{1}{2}\varphi\Delta S
-\frac{i\hbar}{2}\Delta\varphi=0\label{ampl}.
\end{equation}
This is a Fokker-Planck like equation in the
velocity field of the classical action $\nabla S(x,t)$
with a complex diffusion constant $i\hbar/2$.

If the local Schr\"odinger equation around the periodic
orbit has an eigenenergy $E$ the corresponding eigenfunction fulfills
\begin{equation}
\psi_p(t+T_p)=e^{-iET_p/\hbar}\psi_p(t).
\end{equation}
For a general energy value $E$, one can find eigenfunctions
$\psi_p^l(t)$ such that
\begin{equation}
\psi_p^l(t+T_p)=e^{-iET_p/\hbar}\lambda^l_p(E)\psi_p^l(t).\label{loceig}
\end{equation}
If the eigenvalues $\lambda^l_p(E)$ are known the local functional determinant
can be written as
\begin{equation}
\Delta_p(E)=\prod_l(1-\lambda^l_p(E)),
\end{equation}
since $\Delta_p(E)=0$ yields the eigenenergies of the local Schr\"odinger
problem.
We can insert the ansatz (\ref{ansatz})
and reformulate (\ref{loceig}) as
\begin{equation}
e^{\frac{i}{\hbar}S(t+T_p)}\varphi_p^l(t+T_p)=e^{-iET_p/\hbar}\lambda^l_p(E)
e^{\frac{i}{\hbar}S(t)}\varphi_p^l(t).\label{loceig2}
\end{equation}
The phase change is given by the action integral for
one period $S(t+T_p)-S(t)=\int_0^{T_p} L(t)dt$. Using this and the
identity for the reduced action $S_p(E)$ of the periodic orbit
\begin{equation}
S_p(E)=\oint pdq =\int_0^{T_p}L(t)dt+ET_p,
\end{equation}
we get
\begin{equation}
e^{\frac{i}{\hbar}S_p(E)}\varphi_p^l(t+T_p)=\lambda^l_p(E)\varphi_p^l(t).
\label{loceig21}
\end{equation}
Introducing the eigenequation for the amplitude
\begin{equation}
\varphi_p^l(t+T_p)=R^l_p(E)\varphi_p^l(t),\label{loceig3}
\end{equation}
the local spectral determinant can be expressed as
\begin{equation}
\Delta_p(E)=\prod_l(1-R^l_p(E)e^{\frac{i}{\hbar}S_p(E)}).
\end{equation}

At this point we can reexpress the Quantum Gutzwiller-Voros spectral
determinant with the local eigenvalues, which now reads as
\begin{equation}
\Delta(E)=\prod_p\prod_l(1-R^l_p(E)e^{\frac{i}{\hbar}S_p(E)}).
\end{equation}
The trace formula can be recovered from (\ref{7}):
\begin{equation}
\mbox{Tr} G(E)=\frac{1}{i\hbar}\sum_p\sum_l
\frac{R^l_p(E)e^{\frac{i}{\hbar}S_p(E)}}{1-R^l_p(E)e^{\frac{i}{\hbar}S_p(E)}}
\left(T_p(E)-i\hbar\frac{d \log R^l_p(E)}{dE}\right).
\end{equation}
We can also rewrite the denominator as a sum of a geometric series
and get
\begin{equation}
\mbox{Tr} G(E)=\frac{1}{i\hbar}\sum_p\sum_l\left(T_p(E)-i\hbar\frac{d
\log
R^l_p(E)}{dE}\right)\sum_{r=1}^{\infty}(R^l_p(E))^re^{\frac{i}{\hbar}rS_p(E)}.
\end{equation}
The new index $r$ can be interpreted as the repetition number of the
primitive orbit. This expression is the generalization of the
Gutzwiller trace formula for the exact quantum mechanics.
We can see, that the $\hbar$ corrections enter
here in a more complicated form than in the spectral determinant.

The local eigenvalue spectra can be computed with a variety of
numerical methods. These numerical eigenvalues then can be used
to build up the spectral determinant or the generalized trace
formula. For theoretical calculations we develop here an analytic
perturbation method.

\section{Expansion in $\hbar$}

Since $\hbar$ is a small parameter we can develop a perturbation
series for the amplitudes
\begin{equation}
\varphi^{l}(t)=\sum_{m=0}^{\infty} \left(\frac{i\hbar}{2}\right)^m
\varphi^{l(m)}(t)
\label{expa}
\end{equation}
which can be inserted into the equation (\ref{eqs}).
The eigenvalue can also be expanded in powers of $i\hbar/2$:
\begin{equation}
R_l(E)=\exp\left\{\sum_{m=0}^{\infty}
\left(\frac{i\hbar}{2}\right)^m C_l^{(m)}\right\}.
\end{equation}
The $\hbar$ expansion of the eigenvalues is
\begin{eqnarray}
R_l(E)&=&\exp\left\{\sum_{m=0}^{\infty}
\left(\frac{i\hbar}{2}\right)^m C_l^{(m)}\right\}\\
=\exp(C_l^{(0)}) \left( 1 \right. &+&
\frac{i\hbar}{2} C_l^{(1)}+
\left(\frac{i\hbar}{2}\right)^2
\left(\frac{1}{2} (C_l^{(1)})^2+ C_l^{(2)} \right) + ... \; . \label{evexp}
\end{eqnarray}

The eigenvalue equation in $\hbar$ expanded form reads as
\begin{equation}
\sum_{m=0}^{\infty} \left(\frac{i\hbar}{2}\right)^m \varphi^{l(m)}(t+T_p)=
\exp\left\{\sum_{m=0}^{\infty}
\left(\frac{i\hbar}{2}\right)^m C_l^{(m)}\right\}.
\sum_{m=0}^{\infty} \left(\frac{i\hbar}{2}\right)^m \varphi^{l(m)}(t).
\end{equation}
Expanding the eigenvalue like in (\ref{evexp}) and collecting
the terms of the same order in $\hbar$ yield a set of eigenequations
\begin{eqnarray}
\varphi^{l(0)}(t+T_p)&=&\exp(C_l^{(0)})\varphi^{l(0)}(t),\nonumber \\
\varphi^{l(1)}(t+T_p)&=&\exp(C_l^{(0)})[\varphi^{l(1)}(t)+C_l^{(1)}
\varphi^{l(0)}(t)],
\nonumber \\
\varphi^{l(2)}(t+T_p)&=&\exp(C_l^{(0)})[\varphi^{l(2)}(t)+
C_l^{(1)}\varphi^{l(1)}(t)
+(C_l^{(2)}+\frac{1}{2}(C_l^{(2)})^2)\varphi^{l(0)}(t)],\label{1}
\end{eqnarray}
and so on. These equations are the conditions selecting the eigenvectors
and eigenvalues
and they hold for all $t$. Without loss of generality we can also assume
that $\varphi^{l(0)}(0)=1$ and $\varphi^{l(m)}(0)=0$ for $m>0$. By
adding these assumptions we can simplify the equations (\ref{1}):
\begin{eqnarray}
\varphi^{l(0)}(T_p)&=&\exp(C_l^{(0)}),\label{cond} \\
\varphi^{l(1)}(T_p)&=&\exp(C_l^{(0)})C_l^{(1)},
\\
\varphi^{l(2)}(T_p)&=&\exp(C_l^{(0)})(C_l^{(2)}+\frac{1}{2}(C_l^{(2)})^2).
\end{eqnarray}

\section{The analytic eigenbasis}

In the neighborhood of a classical periodic orbit we
can look for the solution of the Hamilton-Jacobi
equation in a power series form.
Let $q(t)$ denote a classical periodic orbit with period
$T_p$.
Let us expand the phase around the time dependent
trajectory:
\begin{equation}
S(x,t)=\sum_n^{\infty} s_n(t) (x-q(t))^{n}/n!,
\end{equation}
where $s_n$ are $d^n$ matrices in dimension $d$ with time dependent
entries.
This power series ansatz assures that the local Schr\"odinger problem
corresponds to the one in (\ref{trg}).
To derive ordinary differential equations for the expansion
coefficients let us expand the potential around the periodic orbit
\begin{equation}
U(x)=\sum_n^{\infty} u_n(t) (x-q(t))^{n}/n!,
\end{equation}
where $u_n$ are $d^n$ matrices in general.
If we put this two expressions in the Hamilton-Jacobi equation
we get in the one dimensional case
\begin{equation}
\dot{s}_n-s_{n+1}\dot{q} + \frac{1}{2} \sum_{l=0}^{n}
\frac{n!}{(n-l)! l!} s_{n-l+1}s_{l+1} + u_n = 0.\label{ampli}
\end{equation}
In the multidimensional case we get similar expressions
for the entries of the $s$ matrices.
If the $s_1$ vector is choosen to be the momentum of the classical orbit
\begin{equation}
p=\dot{q}  = s_1\label{ps1},
\end{equation}
the equations are simpler and their meanings are obvious.
The first eq. in the hierarchy corresponds to the classical
action along the path:
\begin{equation}
\dot{s}_0= \frac{p^2}{2} - u_0 =L(t),
\end{equation}
where $L(t)$ is the Lagrange function evaluated on the periodic orbit.
The second is the Newton equation
\begin{equation}
\dot{p}=-u_1,
\end{equation}
since $u_1$ is the force along the trajectory.
The $d\times d$ matrix $s_2$ is familiar from the wave
packet theory and describes the shape of a gaussian wave packet\cite{Heller}
\begin{equation}
\dot{s}_2=-s_2^2 -u_2.\label{packet}
\end{equation}
$\mbox{Tr} s_2$ describes the expansion of infinitesimal volume elements
evolving along the classical orbit.
The next equation is
\begin{equation}
\dot{s}_3=-3 s_2 s_3 - u_3
\end{equation}
and the rest of the equations are linear equations for $s_n$.
These are pure classical equations describing
the analytic structure of the action around periodic orbits.

The gradient of $S(x,t)$ entering  in the amplitude
equation (\ref{ampl}) are periodic along the periodic orbit.
Therefore, the $s_n, n>0$ matrices are also periodic yielding
the boundary conditions $s_n(t)=s_n(t+T_p)$ where $T_p$ is the period
of the orbit. The term $s_0(t)$ is not periodic and is
given by the action integral $s_0(t)=\int_0^t L(t)dt$.
$s_1(t)$ is the momentum along the periodic orbit and it is a periodic
function.
The most complicated equation is (\ref{packet}). In general it
has more than one periodic solutions. In case of unstable
periodic orbits the solution of the equation (\ref{packet})
converges to a single stable solution starting from almost
all initial conditions. The rest of the solutions are unstable.
The wave packet described by the stable solution is decaying in time,
while the rest of the solutions describe wave packets with
increasing amplitudes. These solutions are non-physical, since
they describe local wave functions with exponentially increasing
norms. We have to exclude these solutions. In case of stable periodic
orbits we also have only one solution of (\ref{packet})
for which the local wave function is decaying and we have to choose
this solution.
The higher order ($n>2$) equations are linear in $s_n$ and their unique
periodic solutions can be found order by order.

\section{Evolution of the amplitude}

After solving locally the Hamilton-Jacobi equation we can look for the
local solution of the amplitude
equation. We can expand the amplitude around the classical path
in power series. This analytic basis is appropriate
for classical Perron-Frobenius operators since it is very easy
to diagonalize the evolution operator on this basis\cite{HHR}.
Inserting the expansion
\begin{equation}
\varphi (x,t)=\sum_n^{\infty} a_n(t) (x-q(t))^{n}/n!
\end{equation}
into the equation (\ref{ampl}) yields the following equations
for the  coefficients in one dimension
\begin{equation}
\dot{a}_n - a_{n+1} \dot{q} +\sum_{l=0}^{n}
\frac{n!}{(n-l)!l!} \left(a_{n-l+1}
s_{l+1}+\frac{1}{2} a_{n-l}s_{l+2}\right) -\frac{i\hbar}{2} a_{n+2}
=0
\label{eqs}.
\end{equation}
In the multidimensional case we get similar equations
for the expansion coefficient matrices.
Using eq. (\ref{ps1}) one can slightly reduce these equations:
\begin{equation}
\dot{a}_0 =- \frac{s_2}{2} a_0 + \frac{i\hbar}{2} a_{2},
\label{equ1}
\end{equation}
\begin{equation}
\dot{a}_1 =- \frac{3s_2}{2} a_1 -
\frac{s_3}{2} a_0+ \frac{i\hbar}{2} a_{3},
\label{equ2}
\end{equation}
\begin{equation}
\dot{a}_2 =- \frac{5s_2}{2} a_2 -
2s_3 a_1 - \frac{s_4}{2} a_0+ \frac{i\hbar}{2} a_{4},
\label{equ3}
\end{equation}
and so on. These equations are linear and have
the general form
\begin{equation}
\dot{a}_n =- \frac{(2n+1)s_2}{2} a_n ... + \frac{i\hbar}{2}
a_{n+2}.
\label{equ4}
\end{equation}
These equations are matrix equations in higher
dimensions but their structure remain unchanged.

\section{Stationary solutions}

The set of equations (\ref{ampli}),(\ref{eqs}) and (\ref{loceig})
define the full set of equations we have to solve. This can be carried
out without
further considerations. However, we know beforehand, that the solutions
of the Scr\"odinger equation are stationary states. The stationarity
condition implies that the phase of the wave function fulfills
the condition
\begin{equation}
\frac{\partial S(x,t)}{\partial t}=-E
\end{equation}
and the amplitude has no explicit time dependence
\begin{equation}
\frac{\partial \varphi (x,t)}{\partial t}=0.
\end{equation}
These equations give us some additional equations for the expansion
coefficients, which have the form
\begin{eqnarray}
s_0(t)-\dot{q}(t)s_{1}(t)&=&-E,\\
s_n(t)-\dot{q}(t)s_{n+1}(t)&=&0 \;\;\;\;\mbox{for $n>0$},\\
a_n(t)-\dot{q}(t)a_{n+1}(t)&=&0 \;\;\;\;\mbox{for $n\geq0$},
\end{eqnarray}
in the one dimensional case.
In the multidimensional case the coefficient matrices fulfill
similar equations.
These equations can help us to reduce the equations we really
have to solve, since some of the higher order expansion coefficients
can be expressed by the time derivatives of the lower order
coefficients. In one dimension all the higher coefficients
can be directly computed from the time derivatives of the
zero order terms. In two dimensions the number of $s_n$ and
$a_n$ matrix elements is $n+1$. The
number of the additional equations derived above is $n$.
Therefore, on each level we get $1$ entirely new equation,
which we have to solve. In three dimensions we get $n$
entirely new equations for the phase and the amplitude
on each level. We show on
examples how the reduction of the equations can be carried
out.

\section{$\hbar$ expansion on the analytic base}

Inserting the expansion (\ref{expa}) in the equations (\ref{eqs}-\ref{equ4})
we get coupled equations. The zeroth order or semiclassical equations
form an autonomous system
\begin{equation}
\dot{a}_n^{(0)} - a_{n+1}^{(0)} \dot{q} +\sum_{l=0}^{n}
\frac{n!}{(n-l)!l!} \left(a_{n-l+1}^{(0)}
s_{l+1}+\frac{1}{2} a_{n-l}^{(0)}s_{l+2}\right)
=0
\label{eq}.
\end{equation}
For example, the first three equations have the form
\begin{equation}
\dot{a}_0^{(0)} =- \frac{s_2}{2} a_0^{(0)},
\label{eq1}
\end{equation}
\begin{equation}
\dot{a}_1^{(0)} =- \frac{3s_2}{2} a_1^{(0)} -
\frac{s_3}{2} a_0^{(0)},
\label{eq2}
\end{equation}
\begin{equation}
\dot{a}_2^{(0)} =- \frac{5s_2}{2} a_2^{(0)} -
2s_3 a_1^{(0)} - \frac{s_4}{2} a_0^{(0)}.
\label{eq3}
\end{equation}
The important feature of these equations is that they are linear and have
the general form
\begin{equation}
\dot{a}_n^{(0)} =- \frac{(2n+1)s_2}{2} a_n^{(0)} + ... ,
\label{eq4}
\end{equation}
and so on.

We can solve equation (\ref{eq1})
and get
\begin{equation}
a_0^{0(0)}(t)=a_0^{0(0)}(0)\exp\left(-\int_0^{t}
\frac{1}{2} s_2(t)dt\right)
\end{equation}
from which one can read off the eigenvalue
\begin{equation}
C_0^{(0)}=-\int_0^{T_p} \frac{1}{2} s_2(t)dt.
\end{equation}
The $s_2$ in general goes through $1/t$ type singularities. If this happens
the integral should be carried out by principal value integration.
This means that we add a regularizer $i\epsilon$ term. The integral then
can be calculated and in the
$\epsilon\rightarrow 0$ limit we recover an $i\pi/2$ Maslov phase.

The rest of the equations do not play a role in yielding the first eigenvalue.
The solution $a_0^{0(0)}(t)$ can
be inserted into the next equation (\ref{eq2}).
Since equation (\ref{eq2}) is a linear one driven by $a_0^{0(0)}(t)$ its
particular solution fulfills the condition (\ref{1}). The rest of the
equations can be solved the same way and we get the eigenamplitudes
$a_n^{0(0)}$. The rest of the semiclassical eigenvalues can be recovered
by setting $a_n^{l(0)}=0$ for $n<l$. Then the $l$-th eigenvalue is
given by
\begin{equation}
C_l^{(0)}=-\frac{2l+1}{2}\int_0^{T_p} s_2(t)dt.
\end{equation}

The semiclassical eigenvalues are connected with the stability
properties of the periodic orbits. For example the
first ($l=0$) eigenvalue is related to the product of
the expanding eigenvalues\cite{Vattay1}
\begin{equation}
\exp(C_0^{(0)})=\frac{e^{i\mu_p\pi}}{|\prod_i \Lambda_i|^{1/2}}
\end{equation}
where $\Lambda_i$ denotes the expanding ($\Lambda_i > 1$) eigenvalues of
the linear stability or Jacobi matrix of the periodic orbit
and $\nu_p$ is the Maslov index of the periodic orbit. The Maslov phase
comes from the singularities of $s_2(t)$ (See e.g. Ref\cite{Vattay1}.)
The product
\begin{equation}
\Delta(E)=\prod_p \prod_l (1 - \exp(iS_p/\hbar+C^{(0)}_l))
\end{equation}
in this approximation is known as the
Selberg product\cite{Selberg,Voros}.

After the calculation of local semiclassical eigenvalues and eigenvectors
we can use them in the next level of approximation.
The differential equations connecting the $m+1$-th order amplitudes
with the $m$-th order amplitudes are
\begin{equation}
\dot{a}_n^{(m+1)} - a_{n+1}^{(m+1)} \dot{q} +\sum_{l=0}^{n}
\frac{n!}{(n-l)!l!} \left(a_{n-l+1}^{(m+1)}
s_{l+1}+\frac{1}{2} a_{n-l}^{(m+1)}s_{l+2}\right) -a_{n+2}^{(m)}
=0
\label{eqsx}.
\end{equation}
Again, using Eq. (\ref{ps1}) one can slightly reduce these equations
\begin{equation}
\dot{a}_0^{(m+1)} =- \frac{s_2}{2} a_0^{(m+1)} +
a_{2}^{(m)},
\label{equx1}
\end{equation}
\begin{equation}
\dot{a}_1^{(m+1)} =- \frac{3s_2}{2} a_1^{(m+1)} -
\frac{s_3}{2} a_0^{(m+1)}+ a_{3}^{(m)},
\label{equx2}
\end{equation}
\begin{equation}
\dot{a}_2^{(m+1)} =- \frac{5s_2}{2} a_2^{(m+1)} -
2s_3 a_1^{(m+1)} - \frac{s_4}{2} a_0^{(m+1)}+ a_{4}^{(m)},
\label{equx3}
\end{equation}
and so on. These equations are linear and have the general form
\begin{equation}
\dot{a}_n^{(m+1)} =- \frac{(2n+1)s_2}{2} a_n^{(m+1)} + ... +
a_{n+2}^{(m)}.
\label{equx4}
\end{equation}
Inserting the eigenamplitudes $a_n^{l(m)}(t)$ we get linear
driven equations for the next order of the amplitudes. The solutions
of these equations, which satisfy the conditions of type (\ref{1}), yield
the corrections $C_l^{(m)}$ of the semiclassical
eigenvalues $C_l^{(0)}$.

It is important to note that the driving terms in
these equations cause resonances, since they are solutions
of the homogenious equations they drive.
The corrections $a_n^{l(m)}(t)$ are not exponentially growing functions
of time. Indeed,
a solution of (\ref{equx1}) behaves like
\begin{equation}
a_n^{l(1)}(r\cdot T_p) \sim \exp(rC_l^{(0)})( a + b\cdot r),
\end{equation}
due to the resonance, where $a$ and $b$ are appropriate constants.
The particular solution satisfying (\ref{cond}) is the one
which does not contain pure exponential part ($a=0$) and
then the correction can be read off as $C^{(1)}_l=b$.

As a consequence of this hierarchy, it is increasingly difficult
to get corrections of the eigenvalues corresponding to large $l$.
It is more convenient to reorganize the quantum Selberg
product as a product of quantum inverse zeta functions
\begin{equation}
\Delta(E)=\prod_l \zeta^{-1}_l(E)
\end{equation}
where the quantum zeta functions are defined by
\begin{equation}
\zeta^{-1}_l(E)=\prod_p\left(
1-\exp( iS_p(E)/\hbar+\sum_m(i\hbar/2)^m C^{p(m)}_l(E)) \right).
\end{equation}
These zeta functions are the quantum generalizations of the
Ruelle zeta functions\cite{Ruelle,AAC}.
The leading resonances and the eigenenergies can be computed
from the zeroes of the $l=0$ quantum zeta function. The curvature
expansion proposed by Cvitanovi\'c and Eckhardt\cite{CE} can
also be applied using the new quantum mechanical weights
\begin{equation}
t_p=\exp\left(iS_p(E)/\hbar+\sum_m (i\hbar/2)^m C^{p(m)}_0(E) \right).
\end{equation}

\section{Billiards}

We have to discuss some special features of the billiard systems
here shortly. In billiards the potential is not an analytic function,
therefore the theory can not be used without further considerations.
We have Dirichlet boundary condition for the wave function
on hard walls. The wave function should vanish on the wall.
Our approach here is basically the tracing of a wave packet along
the classical trajectory in the configuration space. When the packet
is hitting the wall, the incoming wave function at time $t$ is given by
the packet right before the collision, evaluated on the wall
\begin{equation}
\psi_{in}(x(s),y(s),t)=\varphi(x(s),y(s),t_{-0})e^{iS((x(s),y(s),t_{-0})/\hbar},
\end{equation}
where ($x(s),y(s)$) is some analytic parametrization of the wall around
the classical point of reflection.
The outgoing wave function is the wave function right after the
collision
\begin{equation}
\psi_{out}(x(s),y(s),t)=\varphi(x(s),y(s),t_{+0})
e^{iS((x(s),y(s),t_{+0})/\hbar}.
\end{equation}
The sum of the incoming and the outgoing wave functions should vanish
on the hard wall due to the Dirichlet condition.
This implies that the incoming and the outgoing
amplitude and the phase has the relation
\begin{equation}
\varphi(x(s),y(s),t_{-0})=\varphi(x(s),y(s),t_{+0})\label{bou1}
\end{equation}
and
\begin{equation}
S(x(s),y(s),t_{-0})=S(x(s),y(s),t_{+0})+i\pi.\label{bou2}
\end{equation}
These relations mean, that the power series with respect to $s$
of these functions are equal on both sides of the collision modulo
the $\pi$ phase shift. This phase shift can be interpreted as the
Maslov phase coming from the hard wall.

\section{Example I.}

As an example first we study a simple two dimensional potential system
with potential $U(y)=-\frac{1}{2}\lambda^2 y^2-\frac{1}{24}\alpha y^4$,
which is bounded by hard walls in $x=0$ and $x=L$ (Fig. 1). The only classical
periodic orbit in the system is lying on the $y=0$ line and the particle
is moving with constant momentum $p=\sqrt{2E},(m=1)$ along it.

The equations for the $y$ derivatives of the phase are
\begin{eqnarray}
\dot{S}_{yy}&+&S_{yy}^2+U_{yy}=0,\label{as3}\\
\dot{S}_{yyy}&+&3S_{yy}S_{yyy}+U_{yyy}=0,\label{as4}\\
\dot{S}_{yyyy}&+&4S_{yyyy}S_{yy}+3S_{yyy}^2+3S_{yyx}^2+U_{yyyy}=0\label{as7},
\end{eqnarray}
and so on. We will compute here only the first correction, so
we have to solve only these equations.
The stationarity conditions yield
\begin{eqnarray}
\dot{S}_{yy}&-&pS_{yyx}=0,\label{as5}\\
\dot{S}_{y}&-&pS_{yx}=0,\label{as1}\\
\dot{S}_{x}&-&pS_{xx}=0,\label{as2}\\
\dot{S}_{xyy}&-&pS_{xxyy}=0\label{as6}.
\end{eqnarray}

The classical momentum is the first derivative of the phase
\begin{eqnarray}
S_x&=&p,\\
S_y&=&0.
\end{eqnarray}
Since these components are constant in time Eq. (\ref{as1}) and
(\ref{as2}) yield $S_{xy}=0$ and $S_{xx}=0$.
The periodic solutions of (\ref{as3}) are constant in time and
given by $S_{yy}=\pm \lambda$. The stable solution is the one with
positive sign. The periodic solution of (\ref{as4}) is $S_{yyy}=0$.
Since $S_{yy}$ is constant (\ref{as5}) gives us $S_{xyy}=0$.
The solution of (\ref{as7}) then simply $S_{yyyy}=\alpha/4\lambda$.

The zero order amplitude equations for the $y$ derivatives are now
\begin{eqnarray}
\dot{a}&+&\frac{1}{2}S_{yy} a=0,\\
\dot{a}_y&+&\frac{3}{2}S_{yy} a_y+\frac{1}{2}S_{yyy}a=0,\\
\dot{a}_{yy}&+&\frac{5}{2}S_{yy}
a_{yy}+S_{yyx}a_x+2S_{yyy}a_y+\frac{1}{2}(S_{yyxx}+S_{yyyy})a=0.\label{at1}\\
\end{eqnarray}
The solution of the first one is $a=\exp(-\lambda t/2)$. In one
period ($T_p=T=2L/p$) this solution decays as $a(T)=\exp(-\lambda
T_p/2)$ so $C^{(0)}_0=-\lambda T$. Similarly,
$C_l^{(0)}=-(l+1/2)\lambda T_p$. The semiclassical spectral
determinant for this system then reads as
\begin{equation}
\Delta(E)=\prod_l\left(1-\exp(i2pL/\hbar-(l+1/2)\lambda T)\right).
\end{equation}
The first semiclassical zeta function is the term with index $l=0$:
\begin{equation}
\zeta^{-1}_0(E)=\left(1-\exp(i2pL/\hbar-1/2\lambda T)\right).
\end{equation}
Now we compute its first quantum correction. To do this, we have
to solve the first correction equation
\begin{equation}
\dot{a}^{(1)}+\frac{1}{2}S_{yy}
a^{(1)}=a^{(0)}_{yy}+a^{(0)}_{xx}.\label{crr}
\end{equation}
The function $a_{yy}$ is the solution of (\ref{at1}) which decays as
$a_{yy}(nT)=\exp(-\lambda n T_p/2)$ in time.
This solution is
\begin{equation}
a_{yy}(t)=-e^{-\lambda t/2}\frac{\alpha}{16\lambda}.
\end{equation}
The $a_{xx}$ can be calculated using the stationarity relations
\begin{eqnarray}
\dot{a}&-&pa_{x}=0,\\
\dot{a_{x}}&-&pa_{xx}=0.\\
\end{eqnarray}
The result is
\begin{equation}
a_{xx}(t)=-e^{-\lambda t/2}\frac{\lambda^2}{4p^2}.
\end{equation}
The resonant solution of the correction equation (\ref{crr}) is
then
\begin{equation}
a^{(1)}(t)=te^{-\lambda
t/2}\left(\frac{\alpha}{16\lambda}-\frac{\lambda^2}{4p^2}\right).
\end{equation}
{}From here we can read off the first correction as
\begin{equation}
C_0^{(1)}=T\left(\frac{\alpha}{16\lambda}-\frac{\lambda^2}{4p^2}\right).
\end{equation}
The corrected first zeta function then reads
\begin{equation}
\zeta^{-1}_0(E)=\left(1-\exp(i2pL/\hbar-1/2\lambda
T+\frac{i\hbar}{2}\left(\frac{\alpha}{16\lambda}-\frac{\lambda^2}{4p^2}\right)\right).
\end{equation}
This procedure can be continued step by step. From the zeroes of
the first zeta function $\zeta^{-1}_0(E_i)=0$ we can recover the
leading resonances of the system, which are the closest to the real axis
and have the longest lifetime. The appearance of $\alpha$ in the
expression shows, that the first correction is sensitive for
the neighborhood of the periodic orbit.


\section{Example II.}

Our second example is an open billiard system depicted on Figure 2.
The curved wall is locally given by the symmetric curve:
\begin{equation}
x(y)=\frac{1}{2}C_2y^2+\frac{1}{24}C_4y^4+... .
\end{equation}
The separation between the two opposite tips is
$L$. The only periodic orbit is the particle bouncing force and back
between the tips with constant momentum.
This situation occurs for example in the
two disc scattering system
\cite{Gaspard} and in the confocal hyperbolae problem\cite{Niall},
where the first correction to the Gutzwiller trace formula have been
computed numerically.
Since the system is symmetric, it is more convinient to work
on the left part of Fig. 2 and assume, that a vertical
straight hard wall is placed on the symmetry axis.

The expansion coefficients of
the phase and the amplitude fulfill Eqs.(\ref{as3}-\ref{at1}) without
potential term ($U=0$). In addition to these equations we have to
use the bounce conditions (\ref{bou1}) and (\ref{bou2}). If the
classical trajectory is starting from the curved wall, the bounce
conditions give us the following set of equations:
\begin{eqnarray}
S_{yy}(T)+C_2S_{x}(T)&=&S_{yy}(0)+C_2S_{x}(0),\label{boun1}\\
S_{yyy}(T)&=&S_{yyy}(0),\\
S_{yyyy}(T)+6C_2S_{yyx}(T)+C_4S_{x}(T)&=&S_{yyyy}(0)+6C_2S_{yyx}(0)
+C_4S_x(0),\\
a(T)&=&\frac{1}{|\Lambda|^{1/2}}a(0),\\
a_{yy}(T)+C_2a_x(T)&=&\frac{1}{|\Lambda|^{1/2}}(a_{yy}(0)+C_2a_x(0)),
\label{boun2}\\
\end{eqnarray}
where $$|\Lambda|^{1/2}=e^{C_0^{(0)}}.$$

The solutions of the amplitude and phase equations are
\begin{eqnarray}
S_{yy}(t)&=&\frac{1}{t+t_0},\\
S_{yyy}(t)&=&0,\\
S_{yyx}(t)&=&-\frac{1}{S_x(t+t_0)^2},\\
S_{yyyy}(t)&=&\frac{B}{(t+t_0)^4}-\frac{3}{S_x^2(t+t_0)^3},\\
a(t)&=&\frac{t_0^{1/2}}{(t+t_0)^{1/2}},\\
a_x(t)&=&-\frac{t_0^{1/2}}{2S_x(t+t_0)^{3/2}},\\
a_{xx}(t)&=&\frac{3}{4S_x^2(t+t_0)^{5/2}},\\
a_{y}(t)&=&0,\\
a_{yy}(t)&=&\frac{t_0^{1/2}C}{(t+t_0)^{5/2}}+\frac{t_0^{1/2}B}{(t+t_0)^{7/2}}.\\
\end{eqnarray}
The constants $B,C,t_0,\Lambda$ can be set by the bounce conditions
(\ref{boun1}-\ref{boun2}) and we get
\begin{eqnarray}
\Lambda&=&C_2L+1+\sqrt{(C_2L+1)^2-1},\\
t_0&=&\frac{L}{p(\Lambda-1)},\\
B&=&\frac{3L}{p^3\Lambda^2}\frac{1}{(1-1/\Lambda^4)}+
\frac{2C_4L^4}{p^3(\Lambda-1)^4(1-1/\Lambda^4)},\\
C&=&-\frac{pB\Lambda}{2L}\frac{1-1/\Lambda^3}{1+1/\Lambda}-\frac{1}{4p^2}.\\
\end{eqnarray}
The solution of the first correction equation (\ref{crr})
now reads
\begin{equation}
a^{(1)}(t)=(t+t_0)^{-1/2}\int_0^t
dt'(a_{xx}(t')+a_{yy}(t'))(t'+t_0)^{1/2}.
\end{equation}
{}From the solution we can read off the first correction, which is
finally
\begin{equation}
C^{(1)}_0=\frac{1}{p}\left(C_2-\frac{3}{8}C_2
\frac{2+LC_4/3C_2^2}{2+C_2L}\right).\label{result}
\end{equation}
This result can be compared with the findings of ref.\cite{Gaspard}
for the two disk system and with those of ref.\cite{Niall}
for the confocal hyperbolae. In case of the two disk scattering
system $C_2=1/a$ and $C_4=3/a^3$, where $a$ is the radius of the
disk. In this case we get
\begin{equation}
C^{(1)}_0=\frac{5}{8ap},
\end{equation}
which coincides withe the result of Ref.\cite{Gaspard} derived
via Feynman graph technique.
In case of the confocal hyperbolae $C_4=-6C_2^2/L$ and the correction
is
\begin{equation}
C^{(1)}_0=\frac{C_2}{p}.
\end{equation}
This was numerically confirmed in ref.\cite{Niall}.

An important consequence of the result (\ref{result}) is that
in the $L\rightarrow 0$ limit, when the walls are touching each other,
it gives the universal limit
$$C^{(1)}_0=5/8a.$$ In this limit the particles scattering from outside
spend a long time in the horn formed by the touching walls. This can
be considered as the prototype of the intermittent motion.
In this limit the first zeta function in leading order reads as
$$\zeta_0^{-1}(E)=1-\exp\left(i\frac{5}{8ap}\right).$$
The zeroes of this zeta function are
$$\frac{5}{8ap_n} =2\pi (n+1).$$ The energy spectra is given by
$$E_n=\frac{1}{2}\left(\frac{5}{16\pi a(n+1)}\right)^2.$$  We can expect, that
in case of
intermittency due to touching billiard walls, the local energy spectra has an
universal Rydberg type character.

\section{Conclusions}

In this paper we have proposed a new method to evaluate
corrections to the leading saddle point approximation of the
Feynman path integral.
The method reduces the problem to a set of ordinary differential
equations which have to be solved at certain boundary conditions.
In all orders the product structure of the functional determinant
$\Delta(E)$ is maintained. One can introduce the quantum
zeta functions. The corrections to the leading zeta
function is easier to calculate than a general $l>0$ term
and it is very practical to use it for extended computations.
A nice feature of the method is that
the problem of marginally stable periodic orbits can also
be treated, since the differential equations give nonzero
contributions ($C_l\neq 0$) for $l>0$ in case of such an orbit.
The method is applicable
also for general second order partial differential equations, where
the diffusion term is weak. Since the individual saddle
point terms $\mbox{Tr} G_p(E)$ are more accurately
calculated than in the Gauss approximation one can hope that
such a term can describe also the neighborhood of periodic orbits.
This might include effects corresponding to longer periodic
orbits close to reference orbit $p$. One can hope that a few short periodic
orbits with sufficient number of correction terms can accurately
predict energies and resonances.
Numerical calculations using the method outlined in the paper are
in progress.

\section{Acknowledgements}

The author is grateful to E. Bogomolny, P. Cvitanovi\'c, P. Gaspard,
P. Sz\'epfalusy and A. Wirzba for discussions. This work
was partially supported by Phare Accord H 9112-0378, OTKA
2090 and F4286.

\begin{figure}
\vspace{5cm}
\caption{The periodic orbit in the potential $U(y)=-\frac{1}{2}\lambda^2
y^2-\frac{1}{24}\alpha y^4$.}
\end{figure}

\begin{figure}
\vspace{5cm}
\caption{The billiard system of the second example.}
\end{figure}


\begin{thebibliography}{99}
\bibitem{Feynman} R. P. Feynman, Rev. Mod. Phys. {\bf 20},
367 (1948), R. P. Feynman and A. R. Hibbs, Quantum
Mechanics and Path Integrals, McGraw-Hill, New York (1965)
\bibitem{Schulman} L. S. Schulman, {\em Techniques and Applications
of Path Integration}, Wiley-Interscience Publication (1981)
\bibitem{Gutzwiller} M. C. Gutzwiller, J. Math. Phys. {\bf 12},
343 (1971); {\em Chaos in Classical
and Quantum Mechanics} (Springer-Verlag, New York, 1990)
\bibitem{Gaspard} P. Gaspard and D. Alonso, Phys. Rev. {\bf
A47}, R3468 (1993)
\bibitem{Rice} P. Gaspard and S. A. Rice, J. Chem. Phys. {\bf 90}
2225, 2242, 2255 (1989) {\bf 91} E3279 (1989)
\bibitem{Chaos} See examples in : CHAOS {\bf 2} (1) Thematic Issue;
E. Bogomolny and C. Schmit, Nonlinearity {\bf 6}, 523 (1993)
\bibitem{Voros} A. Voros, J. Phys. {\bf A21}, 685 (1988)
\bibitem{Selberg} A. Selberg, J. Indian Math. Soc. {\bf
20}, 47 (1956)
\bibitem{CE} P. Cvitanovi\'c and B. Eckhardt, Phys. Rev. Lett.
{\bf 63}, 823 (1989)
\bibitem{AAC} R. Artuso, E. Aurel and P. Cvitanovi\'c,
Nonlinearity {\bf 3}, 325 (1990)
\bibitem{Wirzba} A. Wirzba, CHAOS {\bf 2}, 77 (1992);
Nucl. Phys. {\bf A560}, 136 (1993)
\bibitem{Berry1} M. V. Berry and K. E. Mount, Rep. Prog. Phys. {\bf 35},
315 (1972)
\bibitem{Buslaev} V. A. Buslaev, Sov. Phys.-Dokl. {\bf 10}, 17-9
(1965)
\bibitem{Ruelle} D. Ruelle, {\em Statistical Mechanics, Thermodynamic
Formalism} (Addison-Wesley, Reading, MA, 1978)
\bibitem{Vattay1} G. Vattay, Prog. Theor. Phys. Suppl. 1993
December Issue, in press
\bibitem{Vattay2} P. Cvitanovi\'c and G. Vattay, Phys. Rev.
Lett. {\bf 71}, 4138 (1993)
\bibitem{Vattay3} G. Vattay, A. Wirzba and P. E. Rosenqvist
preprint, to be published
\bibitem{Vattay4} G. Vattay, A. Wirzba and P. E. Rosenqvist
Proceedings of the ICDC Tokyo 23-26 May 1994, Pergamon Press, in press
\bibitem{Heller} E. J. Heller, J. Chem. Phys. {\bf 62}, 1544 (1975);
{\bf 64}, 63 (1976); W. Eastes and R. Marcus, J. Chem. Phys. {\bf 61},
4301 (1974); W. H. Miller, J. Chem. Phys. {\bf 63}, 996 (1975)
\bibitem{HHR} H. H. Rugh, Nonlinearity {\bf 5}, 1237 (1992);
H. H. Rugh, {\em Thesis}, Copenhagen University (1992);
G. Vattay, {\em Thesis}, E\"otv\"os University Budapest (1992);
P. Cvitanovi\'c, P. E. Rosenqvist, G. Vattay and H. H. Rugh,
CHAOS {\bf 3} (4), 619 (1993)
\bibitem{Keating} M. V. Berry and J. P. Keating, J. Phys.
{\bf A23}, 4839 (1990)
\bibitem{Niall} N. Whelan, CATS preprint, to be published (1994)
\end{thebibliography}
\end{document}